\documentstyle[prd,aps,preprint,epsf]{revtex}
\firstfigfalse
\begin{document}
\draft
\title{ 
 Lattice Gauge Theory Simulations at Nonzero Chemical Potential in the Chiral Limit} 
\vskip -1 truecm
\author{ I. M. Barbour and S. E. Morrison}
\address{Department of Physics and Astronomy, University of Glasgow, Scotland\\
(UKQCD collaboration)}
\author{John~B.~Kogut }
\address {Physics Department, University of Illinois at Urbana-Champaign,
Urbana, IL 61801-30} 
\date{\today}
\maketitle

\begin{abstract}

We present a method of simulating lattice QCD at nonzero chemical potential in the chiral limit.
By adding a weak four-fermi interaction to the standard staggered fermion SU(3) QCD action, we
produce an algorithm in which the limit of massless fermions is well-behaved and physical.
Using configurations at zero chemical potential, and an exact fugacity expansion of the 
fermion determinant, we can simulate QCD at nonzero chemical potential and evade the notorious
problem of the complex action. Small lattice simulations give physical results : At strong gauge
coupling the critical chemical potential $\mu_c$ agrees with theoretical expectations and at weak gauge
coupling $\mu_c$ is nonzero in the low temperature confined phase of QCD and jumps to zero in the
high temperature quark-gluon plasma phase. In all these simulations the quarks are exactly massless
and there is a Goldstone pion.

\end{abstract}

\pacs{ 
12.38.Mh,
12.38.Gc,
11.15.Ha
}
\narrowtext
\section{Introduction}

One of the goals of Lattice Gauge Theory is a thorough investigation of the phase diagram of 
QCD in the temperature-chemical potential plane. Considerable progress has
been made in simulating lattice QCD at nonzero temperature \cite{Karsch}, but there has been little
progress in understanding the theory at nonzero chemical potential $\mu$ even though a sound
lattice formulation was presented twelve years ago \cite{ILCH}. One expects on the
basis of elementary physics arguments \cite{ILCH} that as the chemical potential $\mu$
increases QCD will experience a transition to a quark-gluon plasma at one-third the mass of the 
proton, and that the chiral condensate will serve as an order parameter.
The basic difficulty in simulating QCD at finite density and investigating the
transition quantitatively is that the effective action
resulting from the Grassmann integration over the fermions is {\it complex}
due to the introduction of the chemical potential in the Dirac matrix. 
The standard simulation algorithms \cite{HMC,HMD} for lattice QCD with dynamical fermions
do not apply to this situation. In order to avoid
the determinant, early simulations used the quenched approximation, but they led to an 
unphysical value of the critical chemical
potential $\mu_c$, namely, its proportionality
to the pion mass \cite{DaviesKlepfish}. The physical and mathematical reasons
for this failure have been the subject of considerable debate \cite{LKS}. Recent
work by M. Stephanov \cite{misha}, however, has shown that the quenched model is the $n_f \rightarrow 0$
limit of a theory with both quarks and conjugate quarks, and is not relevant to
QCD with dynamical fermions.

Here we wish to advocate a new approach to simulating QCD with dynamical fermions at nonzero 
chemical potential which has two unique features. First, we approach
the chiral limit of lattice QCD by adding a four-fermi term to its action but setting the bare
masses of the quarks to zero. The four-fermi term is a perturbatively irrelevant interaction 
so it does not have an effect on the values of physical
quantities in the theory's continuum limit. However, in the new algorithm, called $\chi$QCD \cite{LAT96},
the dynamical mass of the quark enters its propagator directly and makes the limit of massless
quarks more accessible, as we will see below. The physics of massless Goldstone pions 
can be studied in $\chi$QCD without the
need for extrapolation. In addition, the algorithm runs one to two orders of magnitude faster than the
traditional one because a physical quantity, the dynamical quark mass rather than the bare lattice
mass, controls the convergence of the inversion of the quark propagator, the most time
consuming step in the procedure. $\chi$QCD approaches the chiral limit of QCD in a different direction
than the standard action and appears to avoid some of its problems.
Second, the fermion determinant is calculated exactly at any $\mu$ and any bare quark mass $m_q$, given
a configuration produced by the $\chi$QCD algorithm at vanishing $\mu$ and any other $m_q$. 
This can be done because of the simple way $m_q$ and $\mu$ enter the theory's action. 
A large ensemble of configurations will generally be needed to produce observables at nonzero $\mu$ 
values with acceptable statistical errors. Preliminary simulations of the method have been
successful and will be presented below.

\section{Formulation}

Consider QCD with a chiral 4-fermion interaction, '$\chi$QCD' \cite{LAT96}.
The molecular dynamics Lagrangian for the lattice theory is 
\begin{eqnarray}
L & = & -\beta\sum_{\Box}[1-\frac{1}{3}{\rm Re}({\rm Tr}_{\Box} UUUU)]
        +\sum_s \dot{\psi}^{\dag} M^{\dag} M\dot{\psi}           \nonumber  \\
  &   & -\sum_{\tilde{s}}\frac{1}{8}N_f\gamma(\sigma^2+\pi^2)      
        +\frac{1}{2}\sum_{\tilde{s}}(\dot{\sigma}^2+\dot{\pi}^2) \nonumber  \\
  &   & +\frac{1}{2}\sum_l(\dot{\theta}_7^2+\dot{\theta}_8^2
        +\dot{\theta}_1^{\ast}\dot{\theta}_1
        +\dot{\theta}_2^{\ast}\dot{\theta}_2
        +\dot{\theta}_3^{\ast}\dot{\theta}_3)
\end{eqnarray}
where 
\begin{equation}
M = \not\!\! D + m_q + \frac{1}{16} \sum_i (\sigma_i+i\epsilon\pi_i)
\end{equation}
and $\beta = 6/g^2$, $\gamma = 1/G^2$,
$\epsilon=(-1)^{x+y+z+t}$, the $\theta_i$ parametrize the SU(3) link variables $U$,
$\tilde{s}$ labels dual sites \cite{CER} on the lattice of volume $V = n_t \times n_s^3$, and 
we have introduced auxiliary fields 
$\sigma$ and $\pi$ to formally linearize the 4-fermi term. 
We use staggered quarks in Eq. 1 because of their relatively good chiral symmetry
properties. For simplicity we consider a theory
where the 4-fermion operator has the 
$$
     U(1) \times U(1)    \subset     SU(N_f) \times SU(N_f)
$$
flavor symmetry generated by $(1,i\gamma_5\xi_5)$ \cite{HKK}. This Lagrangian describes 8 flavors. 
For $N_f$ which is not a multiple of 8 we use ``noisy'' fermions \cite{HMD} and multiply the
fermion kinetic term by $N_f/8$. 

The Dirac operator of standard lattice QCD becomes singular as $m_q \rightarrow
0$. However, inspecting Eq. 2, we see that in $\chi$QCD 
the Dirac operator remains non-singular at $m_q=0$ because 
the auxiliary field $\sigma$ develops a vacuum expectation value
due to chiral symmetry breaking. Conjugate
gradient inversion of the regular QCD Dirac operator requires a number of
iterations which diverges as $V \rightarrow \infty$ and $m_q \rightarrow 0$. Inversion of the
$\chi$QCD Dirac operator requires a finite number of iterations even at
$m_q=0$. 
In addition, the 
scale of the 'time' step in the molecular dynamics algorithm of $\chi$QCD is set by the
dynamical quark mass and can be chosen several times larger here than in the ordinary
lattice QCD algorithm, for the same systematic error \cite{HMD} or acceptance
rate \cite{HMC}.

In order to circumvent the difficulty of investigating the finite-density
transition for dynamical fermions, we build on the method of reference
\cite{BarbourBell}. The method was inspired by the classic work of
Yang and Lee \cite{YL} who showed that the distribution
of the zeros of a partition function determines
the equation of state for a many-body system. For an Ising model in a
magnetic field and for a lattice gas, it was shown that the zeros of the
corresponding partition function lie on a unit circle in the complex
fugacity plane. For a finite system these
zeros will never lie in the physical range for the fugacity, namely
the positive real axis. However, for a temperature below the phase
transition, they will approach the real axis as the volume of the
system tends to infinity. The singularities of the free energy
in the thermodynamical limit are therefore obtained as the
infinite-volume extrapolation of the zeros of the partition
function. A detailed investigation of the finite-size
scaling of the zeros enables one to evaluate
the order of the phase transition and the critical exponent for the
order parameter \cite{Itzykson}. 

This proposed study of lattice QCD at finite density
is based on expanding the Grand Canonical Partition Function
(GCPF)
as a polynomial in the fugacity variable ($e^{\mu n_\tau}$)
where $n_\tau$ is the temporal extent of the lattice. 

The GCPF is given as an ensemble average of the determinant of the Dirac operator
normalised with respect to the determinant at $\mu=0$:
\begin{equation}
Z={{\int [dU][dU^\dagger]\det \hat{M}(\mu)\det M(\mu)e^{-S_g[U,U^\dagger]}}\over {
\int [dU][dU^\dagger]\det \hat{M}(\mu=0)\det M(\mu=0)e^{-S_g[U,U^\dagger]}}}
\end{equation}
The Dirac fermion matrices $M$ and $\hat{M}$ are given by:
\begin{eqnarray}
       2iM_{xy}(\mu) & = & Y_{xy} + G_{xy} + V_{xy} e^{\mu} +V^{\dag}_{xy} e^{-\mu} \nonumber \\
-2i\hat{M}_{xy}(\mu) & = & Y_{xy}^{\dag} + G_{xy} + V_{xy} e^{\mu} +V_{xy}^{\dag} e^{-\mu}.
\end{eqnarray}
where $Y = 2i( m_q + \frac{1}{16} \sum_{<x,\tilde x>} (\sigma( \tilde x)
                   +i\epsilon\pi( \tilde x)))\delta_{xy}, \nonumber$
G contains all the spacelike links and V all the forward timelike links.

The determinant of the Dirac operator is complex (for $\mu\neq 0$) in the $N_c = 3$ case
because $e^{\mu}$ favors forward propagation through Eq. 4 while $e^{-\mu}$ hinders
backward propagation \cite{ILCH}.

The determinants of the fermion matrices $M$ and $\hat{M}$ are related to that of
the propagator matrix \cite{Gibbs}
\begin{equation}
P=\left(\begin{array}{cc}
-GV-YV & V \\
  -V   & 0
\end{array} \right)
\end{equation}
by
$\det(2iM) = e^{N_c\mu n_s^3 n_t} \det(P-e^{-\mu})$ and
$\det(2i\hat{M})=e^{N_c\mu n_s^3 n_t} \det( (P^{-1})^{\dag} - e^{-\mu})$
on an $n_s^3 \times n_t$ lattice.

Since the eigenvalues of $P$ have a $Z(n_t)$ symmetry (since $P$ is proportional to $V$) and because
the complex conjugate configuration is equally probable
in the ensemble average, the GCPF can be expanded as
\begin{equation}
{Z}=\sum_{n=-2N_c{n_s}^3 }^{2N_c{n_s}^3}<b_{|n|}>e^{n\mu {n_t}}
=\sum_{n=-2N_c{n_s}^3}^{2N_c{n_s}^3}e^{-(\epsilon_n - n\mu)/T}.
\end{equation}
The $b_n$ are determined from the eigenvalues of $P^{n_t}$.

For $N_c=3$, the tunneling between the different $Z(3)$ vacua should eliminate
the triality non-zero coefficients. In the simulations described below we
do observe strong cancellations in these coefficients for $n=0$ around zero.

The zeros of this polynomial are the Lee-Yang zeros in the complex
fugacity plane. Their distribution and behaviour with respect to
finite volume scaling will be discussed in a future publication. 

\section{Results}

We consider a description of the system in terms of the canonical partition
functions for fixed particle number\cite{Feynmann}. 
The chemical potential as a function
of the baryon number density
is obtained from the local derivative
of the energy, $\epsilon_n$, of the state with $n$ fermions with respect to $n$.

\begin {equation}
\mu(\rho) = {{1}\over {2 N_c n_s^3 }}{{\partial \epsilon_n}\over{\partial\rho}}
\end {equation}
where $\rho$ is the fermion density, $n\over{2 N_c n_s^3}$.

Although some of the $\epsilon_n$ are determined with large error and have
imaginary part $i\pi$ because the corresponding $b_n$ is negative (in principle,
many more measurements should give all triality zero $b_n$ positive),
it is clear from the simulations that, for all mod(n,3)=0, their real
parts form a continuous curve in $n$, as shown in Fig. 1. 

With this assumption we made a cubic
spline fit to a randomly selected subset of $n_s^3/3$ of the $2n_s^3$ $\epsilon_n$'s 
with triality zero and
evaluated the derivative. This process was performed $n_s^3/3$ 
times and an estimate of the
fitting error and the mean estimated from the distribution 
of the corresponding $\mu(\rho)$ resulted.

\begin{figure}[htb]
\epsfxsize=3.6in
\centerline{\epsffile{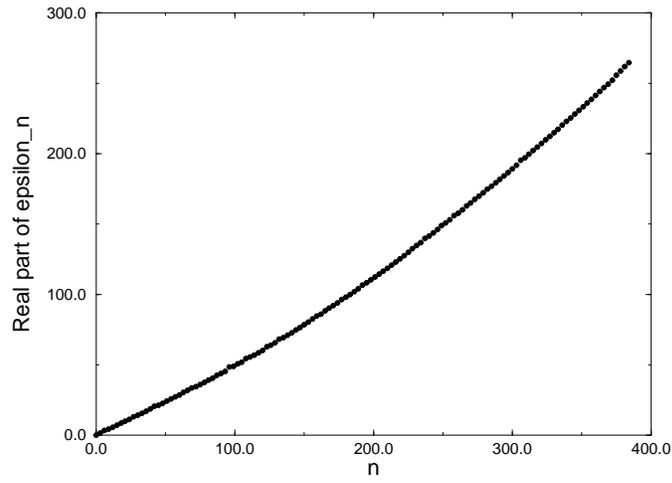}}
\vspace{-0.10in}
\label{fig:6^4_b5_epsilon}
\caption{$\epsilon_n$ vs. $n$, $6^4$ lattice, $\beta=5.0$ and $\gamma=5.0$}  
\vspace{-0.10in}
\end{figure}

\begin{figure}[htb]
\epsfxsize=3.6in
\centerline{\epsffile{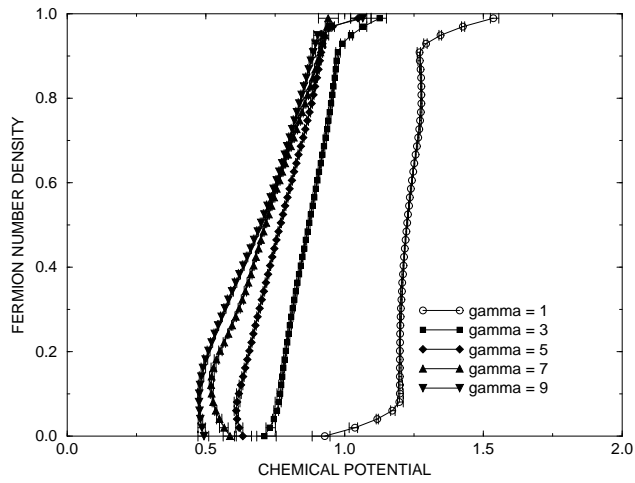}}
\vspace{-0.10in}
\label{fig:4^4_b0.5_density}
\caption{Fermion density vs. chemical potential, $4^4$ lattice, $\beta=0.5$}  
\vspace{-0.10in}
\end{figure}

\begin{figure}[htb]
\epsfxsize=3.6in
\centerline{\epsffile{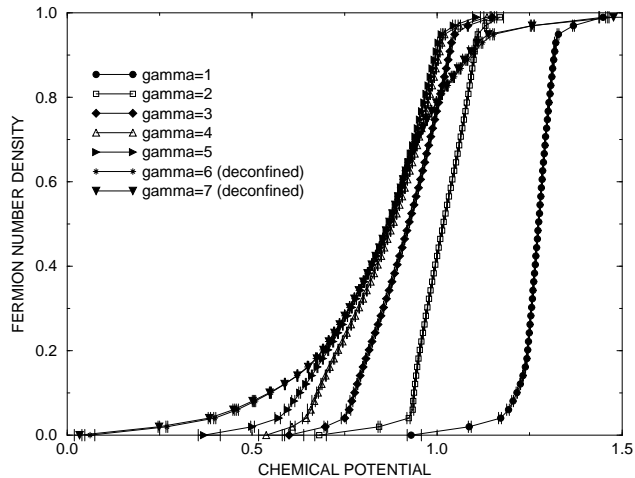}}
\vspace{-0.10in}
\label{fig:6^4_b5_density}
\caption{Fermion density vs. chemical potential, $6^4$ lattice, $\beta=5.0$}  
\vspace{-0.10in}
\end{figure}

Our preliminary simulation results from small lattices are very encouraging. In Fig. 2
we show strong gauge coupling ($\beta= 0.5, m_q=0$) results of the induced fermion number vs. chemical
potential $\mu$ on a $4^4$ lattice at various
four-fermi couplings $\gamma$. Since the pion is strictly massless in these
simulations and since $\mu_c$ is distinctly nonzero for each parameter choice, the
simulation is not afflicted by the diseases of the quenched standard action.
In fact, for large
$\gamma$ where the action reduces to the usual lattice QCD action, mean field
analyses \cite{Bilic}, which should be reliable at strong gauge coupling, predict $\mu_c$ near
0.60, in rough agreement with our figure. 

Next we simulated a weaker gauge coupling in order to approach the continuum limit
of the theory and test the simulation method more strenuously. We chose
$\beta=5.0, m_q=0$ for various $\gamma$ on a larger $6^4$ lattice. Simulations at zero $\mu$
were performed and measurements of the chiral condensate and heavy quark potential indicated a 
finite 'temperature' quark gluon plasma transition between $\gamma$ of 5.0 and 6.0. 
(At $\gamma=5.0$ we measured $<\sigma>=.20$ and the Wilson Line $<WL>=.01$, while at $\gamma=6.0$
we found $<\sigma>=.03$ and $<WL>=.28$.) The task for
our new algorithm was to find this transition and measure the critical chemical potentials
in the 'cold' hadronic phase for $\gamma$ below 5.0 and do the same in 
the 'hot' plasma phase for $\gamma$ above
6.0. Our simulation results, shown in Fig. 3, show this transition very clearly --
for $\gamma$ of 5.0 and smaller the $\mu_c$ is above 0.50, while at $\gamma=6.0$
$\mu_c$ falls abruptly to zero, indicating the presence of unconfined, 'massless'
quarks. Clearly, this estimate of $\mu_c$ is crude. Simulations of simpler models on much larger lattices \cite{3D}
suggests that the 'tails' in the curves in Fig. 3 are finite size effects. 
The finite temperature transition of $\chi$QCD is being studied on larger lattices for 
$\gamma \geq 10$ at variable $\beta$ and more observables are being measured \cite{LAT96} to
obtain quantitative predictions for pure QCD.

Preliminary measurements, on $4^4$ and $6^4$ lattices, of the fermion energy density and condensate using the
stochastic estimator method and including the ratio of the determinants of the Dirac
operator at $\mu$ to that at $\mu=0$ are consistent with the
behaviour of the number density found by the above method. 

At the present time extension to larger lattices is limited by the algorithm used
to determine the eigenvalues of the propagator matrix. The cpu time and storage, to a first
approximation, scale as $n_s^6$ but are independent of $n_t$. The algorithm has been
successfully implemented on an $8^4$ lattice and should be viable on a $10^3 \times n_t$
lattice at its present stage of development.

\section{Conclusions}

Clearly much more work along these lines is called for -- much larger lattices must
be used to extract continuum physics. Although in principle the four fermi term is irrelevant
and the value of $\gamma$ should not effect the values of observables in physical units in the 
continuum limit, practical simulations on finite lattices 
should be done with the four fermi coupling sufficiently small. This 
issue is discussed briefly in \cite{LAT96}, but more quantitative work is needed and is underway. 
Nonetheless, the method has passed several nontrivial tests which we find encouraging.

We are also hopeful that $\chi$QCD can be used
to improve the action of lattice QCD, speed up spectroscopy and
matrix element calculations of conventional hadronic phenomenology, and
determine the universality class of the finite temperature
hadronic matter/quark gluon plasma transition \cite{PW} \cite{KK}.
An algorithm which runs directly in the chiral limit will be very helpful here.

\section{Acknowledgements}

This work was partially supported by NSF under grant NSF-PHY92-00148, 
by PPARC under grant GR/K55554 and by NATO under grant CRG960002.
The simulations were done on the CRAY C-90's at PSC and NERSC. The authors thank
Ely Klepfish, D.K. Sinclair, M. Stephanov, M.-P. Lombardo and A. Koci$\acute{c}$  
for discussions.


\clearpage


\begin{thebibliography}{99}

\bibitem{Karsch} 
K. Kanaya, Nucl. Phys. {\bf B47}, 47 (1996).

\bibitem{ILCH}
J.B. Kogut, H.Matsuoka, M. Stone, H.W. Wyld, S. Shenker, J. Shigemitsu and D.K. Sinclair,
Nucl. Phys. {\bf B225} [FS], 93 (1983). P. Hasenfratz and F. Karsch, 
Phys. Lett. {\bf B125}, 308 (1983).

\bibitem{HMC} 
S. Duane, A.D. Kennedy, B.J. Pendleton and D. Roweth, Phys. Lett. {\bf B195}, 216 (1987).
 
\bibitem {HMD}
S. Duane and J.B. Kogut,  Phys. Rev. Lett. {\bf 55}, 2774 (1985). S. Gottlieb,
W. Liu, D. Toussaint, R.L. Renken and R.L. Sugar, Phys. Rev. {\bf D35},2531 (1987).

\bibitem{DaviesKlepfish}
C.T.H. Davies and E.G. Klepfish, Phys. Lett. {\bf B256}, 68 (1991).

\bibitem{LKS} 
J.B. Kogut, M.-P. Lombardo and D.K. Sinclair. Phys. Rev. {\bf D51}, 1282 (1995); 
Phys. Rev. {\bf D54}, 2303 (1996).

\bibitem{misha}
M. Stephanov, Phys. Rev. Lett. {\bf 76}, 4472(1996).


\bibitem{LAT96}
J.B. Kogut and D.K. Sinclair, hep-lat/9607083.

\bibitem{CER}
Y. Cohen, S. Elitzur and E. Rabinovici, Nucl. Phys. {\bf B220}, 102 (1983).

\bibitem{HKK}
S.~Hands, A.~Koci\'c and J.B. Kogut,
Ann. of Phys. {\bf 224}, 29 (1993) 29.

\bibitem{BarbourBell}
I.M. Barbour and A.J. Bell, Nucl. Phys. {\bf B372}, 385 (1992).

\bibitem{YL}
C.N. Yang and T.D. Lee, Phys. Rev. {\bf 87}, 404 (1952).
T.D. Lee and C.N. Yang, {\it ibid} 410.

\bibitem{Itzykson}
C. Itzykson, R.B. Pearson and J.B. Zuber, Nucl. Phys. {\bf B220}[FS8], 415 (1983).

\bibitem{Gibbs}
P. Gibbs, Phys. Lett. {\bf B182}, 369 (1986)

\bibitem{Feynmann}
For example, see R.P. Feynman, Statistical Mechanics (W.A. Benjamin,1972).

\bibitem{Bilic}
N. Bili$\acute{c}$, K. Demeterfi and B. Petersson, Nucl. Phys. {\bf B377}, 651 (1992).

\bibitem{3D}
S. Hands, S. Kim, and J.B. Kogut, Nucl. Phys. {\bf B442}, 364 (1995).

\bibitem{PW}
R. Pisarski and F. Wilczek, Phys. Rev. {\bf D29}, 
338 (1984).
 
\bibitem{KK}
A. Koci$\acute{c}$ and J. B. Kogut, Phys. Rev. Lett. {\bf 74}, 3109 (1995).

\end{thebibliography}
\end{document}